# Segmentation, Indexing, and Visualization of Extended Instructional Videos

*Alexander Haubold and John R. Kender*
Department of Computer Science
Columbia University
New York, NY 10027
{ah297,jrk}@columbia.edu

**Abstract**

*We present a new method for segmenting, and a new user interface for indexing and visualizing, the semantic content of extended instructional videos. Given a series of key frames from the video, we generate a condensed view of the data by clustering frames according to media type and visual similarities. Using various visual filters, key frames are first assigned a media type (board, class, computer, illustration, podium, and sheet). Key frames of media type board and sheet are then clustered based on contents via an algorithm with near-linear cost. A novel user interface, the result of two user studies, displays related topics using icons linked topologically, allowing users to quickly locate semantically related portions of the video. We analyze the accuracy of the segmentation tool on 17 instructional videos, each of which is from 75 to 150 minutes in duration (a total of 40 hours); the classification accuracy exceeds 96%.*

**Keywords:** instructional video, key frame, segmentation, classification, clustering, indexing, visualization

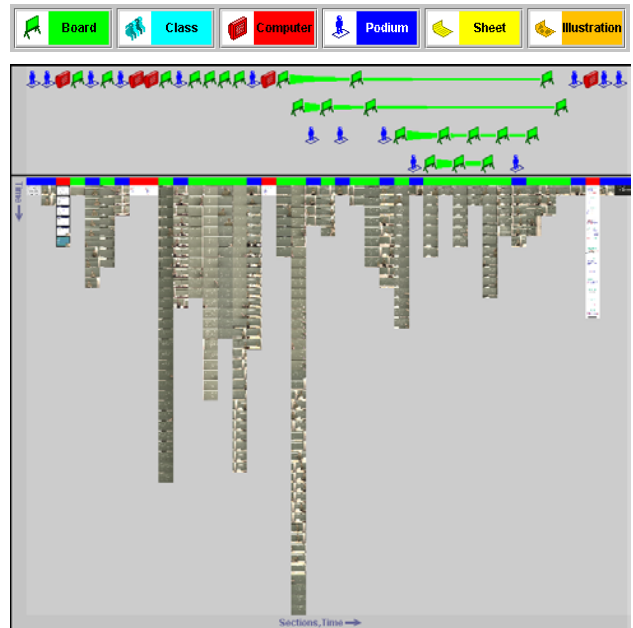

**Figure 1.** Top level of user interface: topological index with key frame summary. (A) above: Each key frame media type is assigned a distinguishable color as well as a descriptive icon. (B) below: Vertical key frame summaries are aligned with media type icons; horizontal topological groupings capture topic commonalities. Icons and key frames are clickable; they select topics, magnify the thumbnails, and pop up the video at the appropriate frame

## 1 Introduction

Video segmentation, indexing, and visualization are essential parts of content-based video retrieval. A successful system would allow the user to review even a long video (up to 150 minutes) by means of some visual summary, and would provide ways to quickly retrieve video contents based on useful clustering and user interface visualizations. Characteristically of their genre, instructional videos tend to be taken in a set environment with a small set of well-defined areas of interest. The segmentation process should exploit this underlying structure.

Research in the area of segmentation and visualization of instructional videos is still in its early stages. Most related work has focused on indexing methods for news videos [1, 2], sports videos [3], and situation comedies [4]. Some work has been done on the segmentation of the blackboard frames of instructional videos taken in a specially instrumented classroom [5]. However, many instructional videos contain material from sources other than the blackboard, and from environments not specifically designed for video analysis.

The summarization and indexing of a video begins by collecting key frame images taken at points of substantial change. In our application, consisting of 17 videos of long lectures, these key frames were chosen by a proprietary software product sensitive to image motion. On average, a key frame is produced every 20 to 25 seconds, so 75 minute lectures contain about 200 key frames, and 150 minute lectures about 350. However, even casual users of these key frames have noted that they tend to be rather repetitive, as much lecturing consists in emphasizing verbally what has been visually created. By grouping together key frames of similar contents into topic clusters, the complexity of the key frame set can be reduced by 80 to 95%.

Structured experiments involving the responses of 11 students and one instructor to three alternative designs yielded a clearer idea of what kind of segmentation, indexing, and visualization would be most useful. The final working design consists of two separate graphs to display the same data by different means (Figure 1): (A) an abstracted Topological View displays the media type and relative (not absolute) temporal location and relationship of topics in the video; (B) a thumbnailed Key Frame View facilitates access to full-size key frames (and the video itself) within each topic. Since key frames from a given topic can appear at any point in the lecture, temporal discontinuities within a topic are illustrated by tapering connecting lines, as seen in Figure 1A. Key frames are further distinguished by their media types. We have identified six: board, class, computer, illustration, podium, and sheet.

These two user interfaces allow the user to search and browse for specific parts of an instructional video, and they also highlight potentially important portions of a lecture. The Topological View automatically lays out interrupted topics in a visually non-interfering planar array; this graphically captures those semantically dense points in the lecture with interaction between different topics. The Key Frame View enables the user to retrieve full-sized key frames in a separate window by sliding the mouse over the thumbnailed images. First (last) key frames for a given topic can be determined by finding the start (end) of the topic using icons in Topological View; clicking on these icons then highlights all the related key frames in the thumbnail strip below, regardless of their temporal separation. (Figure 2)

## 2  Classification by Media Type

The first step in the segmentation process is to assign each key frame to a media type. (Because the media type class occurs so rarely, we have omitted it.) This classification first uses a decision tree of static image feature filters (Figure 4), such as overall color information, color information in certain spatial arrangements, color patterns, and features such as edge information. A post-processing dynamic filter stage follows.

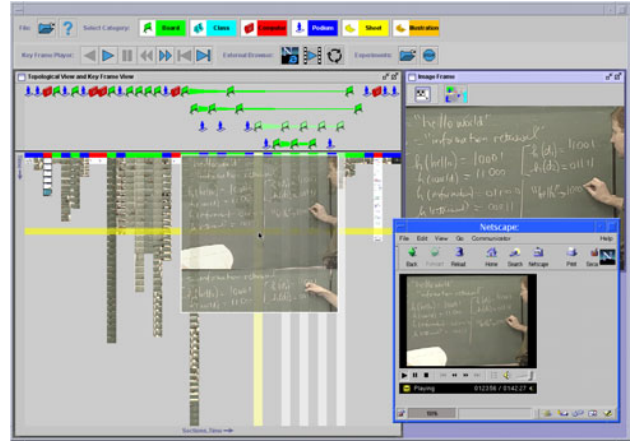

**Figure 2.** User Interface: White vertical bars highlight a selected topic in Key Frame View. Magnified semi-transparent images appear above the thumbnailed key frames, while the key frame summary is browsed with the mouse cursor; a yellow row and column mark the selected thumbnail. Upon clicking on a thumbnail, the full-sized key frame appears in the Image Frame, and the video can be played back in a web browser starting at that key frame. Key frames from a selected topic can also be played back continuously with the Key Frame Player.

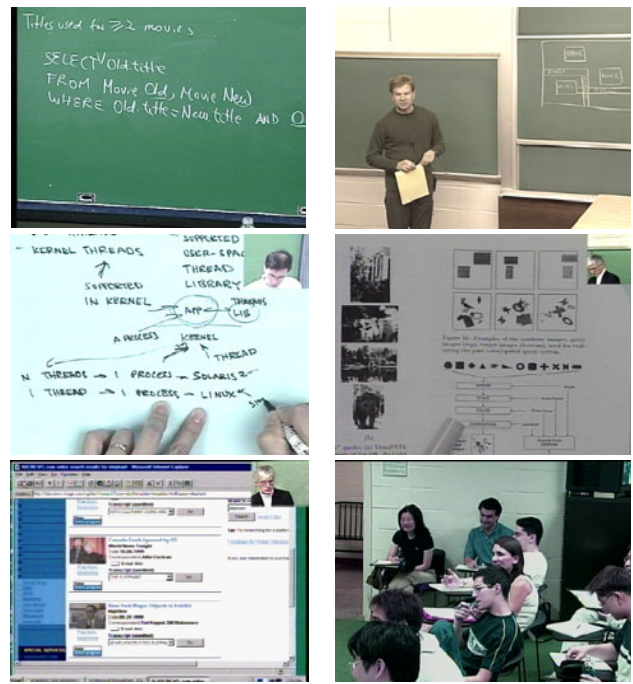

**Figure 3.** Examples from each of the six media types collected from videos of five courses: board, podium, sheet, illustration, computer, class.

### 2.1  Preliminary File Name-Based Classification

In our application, the proprietary key frame detector is aware of two special sources of imagery: the control room-generated title key frames, and PowerPoint key frames. These are stored with distinguished names and are easy to recognize and classify.

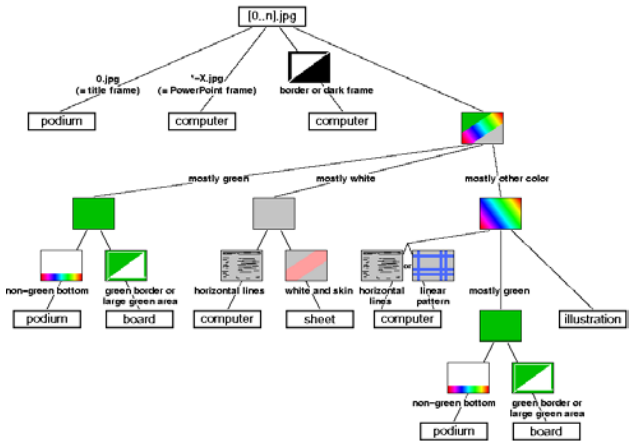

**Figure 4.** Visual feature filters for classification by media type.

### 2.2 First Level Visual Classification

The remaining key frames are first analyzed for characteristic visual features intrinsic to computer-generated material. Empirically it is observed that computer screen images, when scaled to fit the video frame, are padded with a border of black pixels from 5 to 10% of the image. Additionally, because of the need for high ambient lighting for good visual images of the instructor, board, or students, computer key frames are among the darkest key frames. Classifying black bordered or dark frames as computer frames was found to be 100% accurate.

### 2.3 Second Level Visual Classification

Remaining key frames (which may include some additional computer frames) are analyzed by color content. Key frames with mostly green color are labeled as candidates for media types board and podium; those with mostly white color are candidates for computer and sheet; and any remaining are candidates for a more complex analysis for four media types.

#### 2.3.1 Board versus Podium

The semantic difference between the board and podium media types is derived from the behavior of the instructor. When the instructor is using the blackboard, the cameraman tends to focus on the blackboard, to capture blackboard content. If the instructor is interacting with the class, the camera tends to focus on the instructor, to capture gestures and facial expressions. Classification can therefore be based on the spatial arrangement of the blackboard color relative to the instructor.

Podium key frames contain portions of green color concentrated in vertically central regions of the image, with bottom portions of the image colored differently (Figure 3b). Predominantly green key frames lacking green color in their bottom 10% are classified as podium. Although the distinction between podium key frames and board key frames is not crisp, any errors are not critical; the primary consequence is that podium key frames are simply not clustered into topics.

Board key frames empirically are observed to be of two major kinds. The first is a large green area that includes a green lower border. The second is a smaller green area (due to occlusion by the instructor) that nevertheless has a nearly complete green border on all sides.

Board and podium key frames are accurately classified at this stage 97% of the time. Errors consist of predominantly green computer or illustration key frames; some of these are detected and corrected in a post-processing phase (Section 2.4).

#### 2.3.2 Computer versus Sheet

Candidates for the computer media type, which at this stage are primarily white, are distinguished by the presence of the horizontal linearity of their contents: rectangular imagery, tables, menu bars, etc. By using a Laplacian edge detector, we extract an edge image, and compute from it a weighted measure of the presence of horizontal lines, giving exponentially more weight to longer lines. Key frames with a measure above a threshold are classified as computer frames, with 99.9% accuracy. The two frames classified in error were due to the rapid motion of the instructor's pen; this motion generated the digitization interlacing artifact of every other scan line being dark, giving the appearance of horizontal lines.

Remaining (and still primarily white) candidates are classified as sheet media type if they include a sufficiently large amount of white, light gray, and/or skin tones, with 100% accuracy.

#### 2.3.3 Remaining Key Frames

Key frames without dominant green or white regions can still be classified as any media type other than sheet. This last stage of processing handles, for example, zoomed-out frames of the board or podium, or computer frames and illustrations having colors other than white or green. An empirically derived sequence of tests revisits the computer, podium, and board media types, leaving the illustration media type to be the default classification if the other three types fail.

The key frame is first tested for media type computer by computing its horizontal line measure, and a related measure of vertical or horizontal color repetition. Computer frames are not affected by classroom ambient lighting conditions, and are therefore characterized by long vertical or horizontal runs of very similar pixel values. Frames exceeding thresholds in either measure are classified as computer. The majority of classification errors occur at this point; these tests appear to be overly general, although all remaining computer frames are classified 100% accurately.

Frames not meeting these conditions are reexamined for the heuristic features specified for board and podium given in Section 2.3.1, with similar classification accuracy.

Failure of these tests results in the key frame being classified as media type illustration. Illustration key frames are mislabeled as computer frames about 23% of the time. Illustrations are typically extracted from printed media that exhibit horizontal linearity and, to some extent, color repetition.

*2.4 Post Processing*

The heuristics in Section 2.3 are based on static image features. Two post-processing methods exploit temporal relationships to verify and refine the media type classification.

The first post-processing method focuses on the small number of fade-out key frames at the end of a video that mirror the fade-in at its beginning. The video begins and ends with the same black panel with white lettering that identifies the lecture and instructor, generated by the control room's computer. They should be classified as podium frames, but may have been misclassified, depending on the content they have been dissolved with. These frames are therefore reexamined for large dark areas, in a temporally reverse manner (Figure 5).

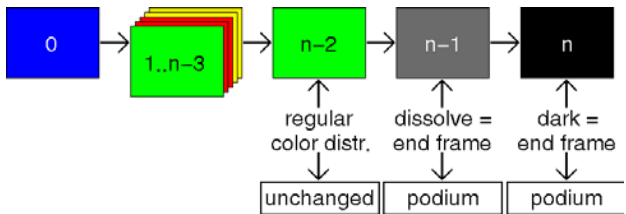

**Figure 5.** Decision graph for re-classifying end key frames.

A second post-processing method notes that a sequence of computer frames might include key frames with predominantly green color; these may have been misclassified as board or podium. The beginning and end of a computer subsequence are easily and reliably detected. The color of those key frames classified as board or podium frames between these endpoints are compared to their computer frame neighbors, and reclassified as computer frames if similar enough. This heuristic corrects about 2/3 of these errors.

## 3   Topological Segmentation

The proprietary software that selects key frames does so in way that is mostly sensitive to instructor motion, concentrating its captures during periods of relative visual calm. Consequently, the key frames are highly redundant. Nevertheless, these frames provide a good measure on how information in the classroom is developed and altered, and illuminate differences in teaching styles: some instructors tend to distribute information linearly, while others like to jump back and forth between topics.

In order to retain these characteristics while reducing redundancy, we cluster similar key frames into topics based on a key frame's visual content. Two key frames are clustered together if the more recent frame elaborates on the visual information found in the more distant one. This clustering reduces a large number of key frames to a set of clusters (topics) of similar key frames that is 5 to 20% of that size. An example is given in Figure 6. Assigning a letter to each key frame denoting its topic, a long string of similarly labeled frames results.

$X^1 \ X^2 \ Y^6 \ \mathbf{A}^1 \ X^{10} \ \mathbf{B}^8 \ X^3 \ Y^1 \ Y^1 \ \mathbf{C}^{29} \ X^{12} \ \mathbf{D}^{11} \ \mathbf{E}^{21} \ \mathbf{F}^{15} \ \mathbf{G}^{28} \ X^{16} \ Y^1$
$\mathbf{H}^7 \ \mathbf{I}^{42} \ X^4 \ \mathbf{I}^5 \ X^1 \ \mathbf{H}^2 \ \mathbf{I}^8 \ X^{10} \ \mathbf{J}^{14} \ X^1 \ \mathbf{K}^7 \ \mathbf{J}^1 \ \mathbf{K}^6 \ \mathbf{J}^2 \ \mathbf{K}^{11} \ \mathbf{J}^5 \ X^6 \ \mathbf{J}^5$
$\mathbf{H}^3 \ \mathbf{I}^1 \ X^1 \ Y^{13} \ X^1 \ X^1$

**Figure 6.** For the video displayed in Figure 1, 323 key frames have been reduced to 41 clusters. *X* and *Y* denote podium and computer key frames, respectively. Letters **A** though **K** denote distinct clusters of board key frames. The exponent for each letter denotes how many frames of the same topic appear in each contiguous sequence.

We apply topological clustering only to board and sheet key frames, because these are the most informative media types. (It is unclear how computer, podium, or class frames can be clustered according to visual contents: computer frames are often not elaborated but simply superceded, and podium and class frames have no distinctive visual content.) The clustering of board and sheet key frames is broken into two steps. First, images are filtered to extract writing. Secondly, images of this writing are matched using selected sub-windows of content.

*3.1 Filtering*

The essential distinctions between media types has an operational consequence: different sets of filters are applied to either type. For board key frames, this means filtering out the board and all surrounding artifacts, while sharpening the chalk marks. For sheet key frames, a filter removes all areas that do not relate to the material written on the lighter background.

The board filter addresses the problem of poor board versus chalk contrast; Figure 7B is an example of low contrast. The board filter explicitly encodes algorithms to separately extract the background and the foreground. Each step is represented by a branch of the flow diagram in Figure 7A: background in (b) through (e); foreground in (f) through (i); their merger in (j) and (k).

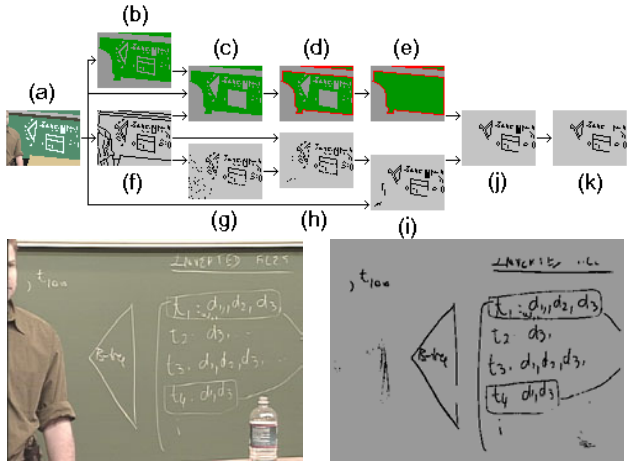
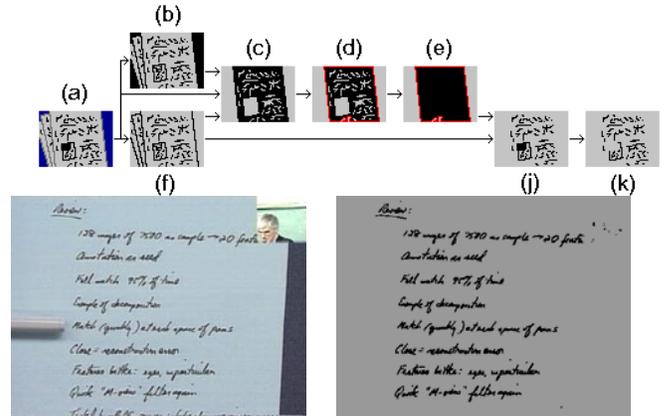

**Figure 7.** Details of the board media type filters. (A) above: flow diagram of filtering process. (a) original image, (b) green color filter, (c) flooded board, (d) outline of flooded area, (e) complete flooding of outlined area (board), (f) edge filter, (g) horizontal and vertical color similarity filter, (h,i) morphological filters, (j) ANDed combination of (e) and (i) results in extraction of board contents, (k) large pixel blob filter removes useless features; (B) below: example of video frame (left) and result of filtering (right).

For board frames, the background is extracted in 4 steps. First, potential blackboard pixels are isolated with a simple green color filter 7A(b). Using the edges found by the edge filter in 7A(f) and the potential blackboard pixels from 7A(b), the board in 7A(a) is flooded to obtain the largest closed blackboard region(s) in 7A(c). Because the result excludes any writing, the result is outlined in 7A(d) and flooded again in 7A(e).

The foreground is extracted beginning with a 3x3 Laplacian edge filter in 7A(f). Edge artifacts from the borders of homogeneous regions are detected by a horizontal and vertical color similarity filter and removed in 7A(g). A morphological filter in 7A(h) removes noise while restoring content pixels mistakenly removed from 7A(f). A second morphological filter in 7A(i) restores content pixels mistakenly removed from 7A(a). The foreground pixels are ANDed with background pixels to recover writing pixels on the board. In a final step 7A(k), large blobs of writing are removed from the filtered image because they are not useful in matching. We call this final binary frame the derived content frame.

The sheet filter uses the same methodology for extracting writing from sheets of paper. However, because ink on fresh sheets of paper usually has higher contrast than chalk on erased blackboards, no noise filtering is necessary for the foreground. The background is extracted identically, using a simple white color filter in 8A(b), a flood fill in 8A(c), an outline of the sheet of paper in 8A(d), and a final flood fill in 8A(e). The edges found in 8A(f) are then directly ANDed with the background pixels in 8A(j), and large blobs of writing are removed in 8A(k), the derived content frame.

**Figure 8.** Details of the sheet media type filter: (A) above: flow diagram of filtering process, (a) original image, (b) white color filter, (c) flooded sheet, (d) outline of flooded area, (e) complete flooding of outlined area (sheet), (f) edge filter, (j) ANDed combination of (e) and (f) results in extraction of sheet contents, (k) large pixel blob filter removes useless features; (B) below: example of input image (left) and output image (right)

### 3.2 Matching

Matching the content between two key frames is complicated by the three degrees of freedom allowed to the otherwise fixed cameras: tilt, pan, and zoom. These introduce translation, scale, and perspective changes. Perspective is handled by an implicit para-perspective method: both frames are considered to be made up of small local windows. As few scale-invariant features are expected, scale is explicitly modeled by successively rescaling one of the pair by a range of 14 scaling factors (from 0.6 to 1.7), experimentally derived. Therefore, starting at full scale, expansions and contractions are alternated until a match is found, or the range is exhausted.

#### 3.2.1 Matching two Key Frames

Given two key frames $i$ and $j$, with $j$ the more recent, we extract a set of features in $i$ by means of an interest operator, and find their correspondences in $j$. If sufficient similarity exists, frame $j$ is considered to be an elaboration of the topic in frame $i$.

Features in frame $i$ are extracted in the form of up to 6 identically sized, wide aspect ratio sub-windows of interest (Figure 9). The selection of sub-windows proceeds as follows:

1. Divide the derived content frame into three equal vertical strips. (Figure 9A(c))
2. For each strip, scan the wide aspect ratio interest window over a coarse grid of locations in the image. (Figure 9A(b,c))
3. Count the content pixels $cc$ in this window placement. If $low \leq cc \leq high$, stop and report window position. (Figure 9A(d))

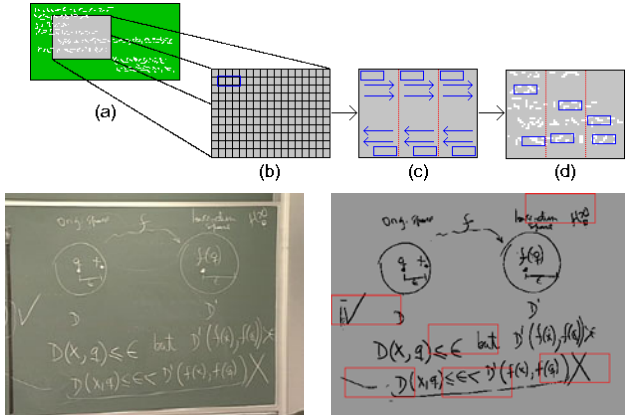

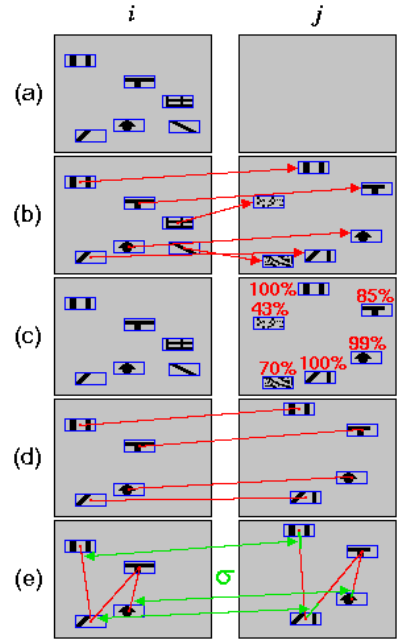

**Figure 9.** (A) above: (a) A board key frame may only contain a portion of the full blackboard. (b) Content pixels from the filtered image are extracted from fixed position in a superimposed grid. (c,d) Up to 6 interesting sub-windows are identified in 3 equally sized vertical strips over the grid. (B) below: example of key frame (left) and 6 most interesting windows (right)

4. Repeat 2 and 3, except scanning bottom-up.
5. If both reports are the same, keep only one.

This heuristic search reflects the empirical observations of both the units of writing and the camera motions observed in the videos. Since panning for boards dominates tilting for sheets, the search enforces and favors a horizontally balanced window acquisition, but without neglecting vertical distribution. Windows are sized so that their height roughly corresponds to two lines of text, while their width is about twice that size to reflect the lengths of average words.

Empirically, it is observed that windows of interest ideally contain between 5 to 30% content pixels. This range provides enough pixels to match with, but not so many as to prohibit the creation of distinctive configurations of pixels .

Having found windows in frame $i$ (Figure 10(a)), to increase match likelihood we next blur the derived content frame $j$ slightly by opening it with a 3x3 mask. We now find the best correspondences to these windows in the blurred derived content frame $j$, and compute a total match score as follows (Figure 10):

1. Find the best location of each sub-window from $i$ in $j$ in the usual manner of template matching (10(b)).
2. Define image match quality for each sub-window match as the amount of matching writing divided by the total amount of writing (10(c)).
3. Define consistency of translation of the windows as the negative of the standard deviation of the lengths of the translation vectors for each sub-window pair (10(d)).

**Figure 10.** Matching algorithm. (a) find interesting sub-windows in i; (b) find corresponding sub-windows in j; (c) determine image match quality; (d) compute σ of translation vectors; (e) compute σ of change of intra-window distances

4. Define consistency of spatial arrangement as the negative of the standard deviation of the errors between corresponding intra-window distances (10(e)).
5. Total match score is a weighted sum of the number of windows in the match, image match quality, translation consistency, and spatial arrangement consistency.

### 3.2.2 Matching all Key Frames

We define a topic $T_k$ to be a temporally ordered but possibly non-consecutive sequence of key frames. Topics are themselves temporally ordered by the time of their most recent frame, $f_k$. The key frames of the video can now be clustered into topics by having each successive key frame either extend an existing topic sequence or start a new one:

1. The first key frame of the video forms the first topic.
2. Each succeeding key frame of the video is matched to the most recent frame of the most recent topic.
3. If this match succeeds, the most recent topic is extended by the incoming frame, and the frame becomes the most recent frame of the topic (Figure 11, case 1).

| Identified topics: | $T_1 = \{ f_1, f_3, f_4 \}$ |
| --- | --- |
| | $T_2 = \{ f_2, f_5, f_7 \}$ |
| | $T_3 = \{ f_6, f_8 \}$ |
| Sequence of topics with most recent topic as last element: | $T = \{ T_1, T_2, T_3 \}$ |
| Next analyzed key frame: | $f_9$ |

| Possible cases | New sequence for $T_k$ and $T$ |
| --- | --- |
| 1. $f_9$ matches $T_3$ | $T_3 = \{ f_6, f_8, f_9 \}$ |
| | $T = \{ T_1, T_2, T_3 \}$ |
| 2. $f_9$ matches $T_1$ | $T_1 = \{ f_1, f_3, f_4, f_9 \}$ |
| | $T = \{ T_2, T_3, T_1 \}$ |
| 3. $f_9$ does not match any $T_k$ | $T_4 = \{ f_9 \}$ |
| | $T = \{ T_1, T_2, T_3, T_4 \}$ |

**Figure 11.** Example of matching heuristics outlining the three cases for a match.

4. If the match fails, the incoming key frame is matched in sequence to the most recent frame of the other topics, in the order of topic recency.
5. If the incoming key frame finds a match, it extends that topic, and it becomes the most recent frame of the topic, and the topic becomes the most recent topic (Figure 11, case 2). If no match is found at all, the incoming key frame starts a new topic, and this new topic becomes the most recent topic (Figure 11, case 3).

Because for media type sheet there are no erasures, the matching is performed exactly as specified in Figure 12, by finding sub-windows in the older frame and searching for them in the newer one. For media type board, where erasures are common and temporal order does not guarantee increased content, the matching is generalized and can also proceed in the reverse direction as well.

## 4 Analysis of Cost of Matching

We now show that, under reasonable assumptions, the matching algorithm is approximately linear. We assume that the number of topics grows linearly but slowly; as seen in the actual data. More specifically, in the 17 videos the ratio of topics to total number of key frames is small, with average = 14.8 / 200 = 0.074. Statistically, a match between two consecutive key frames occurs 89% of the time, while a match between two non-consecutive key frames occurs 3.6% of the time, and no match occurs (i.e. a new topic is formed) 7.4% of the time.

We have calculate the cost of matching as follows:

The expected cost of a match at frame $f$ is composed of three terms: the expected cost of performing a match with the previous key frame (which happens always), plus the expected cost of finding a match with a key frame from a prior topic (which happens when a prior topic is extended), plus the expected cost of performing a match with a key frame from all prior topics (which happens when a new topic is started):

$$Match(f) = p_{exact} * 1 + p_{previous} * \frac{O(f)}{2} + p_{new\ topic} * O(f)$$

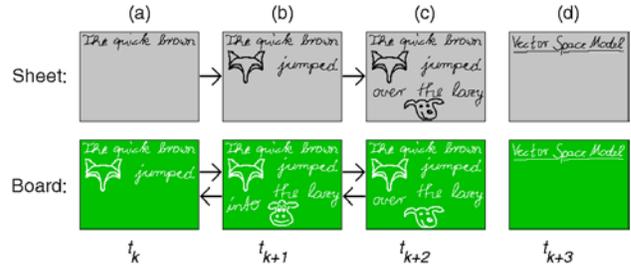

**Figure 12.** Development of contents on media types sheet and board. (a,b,c) Sheet contents are developed strictly forward, so matching is likewise strictly forward. (d) A new topic is easily recognized by no continued sub-windows. In contrast, board contents are subject to erasure (b,c), so matching is two-way. But new topics are still easily recognized (d).

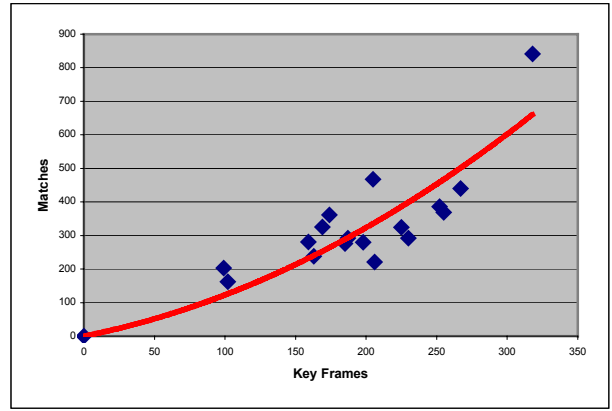

**Figure 13.** Quadratic regression of number of matches versus number of key frames over 17 different instructional videos: $M(f) = .84f + .0039f^2$.

The term $\frac{O(f)}{2}$ signifies the average search space of half the topics for matching a key frame to a continuing topic.

Total Cost: $M(f) = \sum_{f = frames} Match(f)$

Empirically, there are an average of 200 key frames per board or sheet segmentation; the probability of matching the current topic is 178/200 = .89; the probability of matching with a topic prior to the most recent topic is 7.2 / 200 = 0.036; and the probability of not matching with any previous topic is 14.8 / 200 = 0.074. The cost for matching 200 frames becomes:

$$M(f) = \sum_{1}^{200} .89*1 + .036*\frac{O(f)}{2} + .074*O(f)$$

Using $O(f) = f * \frac{14.8}{200} = f * .074$, we compute:

$$M(f) = \sum_{1}^{200} .89*1 + .0013f + .0055f = \sum_{1}^{200} .89 + .0068f$$

$$M(f) = .89f + .0034f^2$$

This result is very close to the quadratic regression in Figure 13, where $M(f) = .84f + .0039f^2$. For most videos, the quadratic term is negligible.

## 5   Accuracy of Clustering Results

While no attempt has been made to optimize the performance of the segmentation tool, we have collected data over 17 extended videos measuring 40 total hours that suggests several properties of the underlying processes.

Media type classification is rather robust, as most media types were correctly detected between 97 and 100% of the time (Table 1). The only exception is detection of type illustration. Topological Segmentation performed equally well at a success rate of more than 96% (Table 2). On a per video basis, the number of topics in error is less than one.

## 6   Conclusion

We have developed and evaluated a novel way of segmenting and visualizing long instructional videos based on key frames (200 to 350). By classifying frames by media type, and clustering hand-written frames by topic, the complexity of browsing this large number is reduced by 80 to 95%. Experiments with 17 instructional videos have shown that the processes enjoy a high (> 96%) success rate. Moreover, we have modeled the cost of the algorithm and confirmed empirically that it is almost linear.

We have demonstrated a novel user interface that visualizes this segmentation by media type and clustering by topic for a given instructional video. Users are able to browse through the lecture's contents by making use of an abstracted icon-based Topological View and a thumbnailed Key Frame View. Experiments with students have shown the interface is useful in quickly gaining an understanding of a video's structure and locating specific portions of a lecture.

Future investigations would include more careful optimizations of the match, particularly with regard to scale, and the automatic extraction of significant terms or diagrams from repeated key frames to augment the Topological View with symbolic index terms.

| N = 4479 | | Classified | | | | |
|---|---|---|---|---|---|---|
| | | B | P | S | I | C |
| Actual | B (696) | - | 0 | 0 | 0 | 0 |
| | P (431) | 0 | - | 2 | 9 | 0 |
| | S (2708) | 0 | 2 | - | 0 | 2 |
| | I (387) | 0 | 0 | 0 | - | 89 |
| | C (257) | 0 | 9 | 0 | 0 | - |

**Table 1.** Confusion Matrix for (B)oard, (P)odium, (S)heet, (I)llustration, (C)omputer: number of key frames that were incorrectly classified. Except for the 23% error of I→C, errors are very small (less than 3%).

| Key Frames | Key Frame Match Errors | Topics Actual | Topics Found |
|---|---|---|---|
| 3394 | 112 | 239 | 252 |

**Table 2.** Overall performance of topological clustering for 17 videos. Errors in matching include mismatches and incorrect starts of new topics.

## Acknowledgements

This work was supported in part by NSF grant EIA-00-71954 and an equipment grant from the Microsoft Corporation.